\documentclass[twocolumn]{pasj00}
\usepackage[dvips]{color}

\begin{document}
\SetRunningHead{N. Narita et al.}{JHK$_{\rm s}$
Simultaneous  Transit  Photometry  of  GJ1214b}
\Received{2012/08/09}
\Accepted{2012/10/11}

\title{IRSF SIRIUS \textit{JHK}$_{\rm s}$ Simultaneous Transit Photometry of GJ1214b}

\author{
Norio \textsc{Narita},\altaffilmark{1,7}
Takahiro \textsc{Nagayama},\altaffilmark{2} 
Takuya \textsc{Suenaga},\altaffilmark{3}
Akihiko \textsc{Fukui},\altaffilmark{4}\\
Masahiro \textsc{Ikoma},\altaffilmark{5}
Yasushi \textsc{Nakajima},\altaffilmark{1,6}
Shogo \textsc{Nishiyama},\altaffilmark{1}
and Motohide \textsc{Tamura}\altaffilmark{1}
}

\altaffiltext{1}{
National Astronomical Observatory of Japan,
2-21-1 Osawa, Mitaka, Tokyo, 181-8588, Japan
}

\altaffiltext{2}{
Department of Astrophysics, Nagoya University,
Furo-cho, Chikusa-ku, Nagoya, 464-8602, Japan
}

\altaffiltext{3}{
Department of Astronomy,
Graduate University for Advanced Studies,
2-21-1 Osawa, Mitaka, Tokyo 181-8588, Japan
}

\altaffiltext{4}{
Okayama Astrophysical Observatory, National Astronomical Observatory of Japan,\\
3037-5 Honjo, Kamogata, Asakuchi, Okayama 719-0232, Japan
}

\altaffiltext{5}{
Department of Earth and Planetary Science, The University of Tokyo,
7-3-1 Hongo, Bunkyo-ku, Tokyo, 113-0033, Japan
}

\altaffiltext{6}{
Hitotsubashi University
2-1, Naka, Kunitachi, Tokyo, Japan, 186-8601
}

\altaffiltext{7}{
NAOJ Fellow
}
\email{norio.narita@nao.ac.jp}


\maketitle

\begin{abstract}
We report high precision transit photometry of GJ1214b
in JHK$_{\rm s}$ bands taken simultaneously with the SIRIUS camera
on the IRSF 1.4~m telescope at Sutherland, South Africa.
Our MCMC analyses show that the observed planet-to-star radius ratios
in JHK$_{\rm s}$ bands are
$R_{\rm p}/R_{\rm s,J} = 0.11833 \pm 0.00077$,
$R_{\rm p}/R_{\rm s,H} = 0.11522 \pm 0.00079$,
$R_{\rm p}/R_{\rm s,Ks} = 0.11459 \pm 0.00099$, respectively.
The radius ratios are well consistent with the previous studies by
\citet{2011ApJ...743...92B} within 1$\sigma$,
while our result in K$_{\rm s}$ band is shallower than and
inconsistent at 4$\sigma$ level with the previous measurements
in the same band by \citet{2011ApJ...736...78C}.
We have no good explanation for this discrepancy at this point.
Our overall results support a flat transmission spectrum
in the observed bands,
which can be explained by a water-dominated atmosphere or
an atmosphere with extensive high-altitude clouds or haze.
To solve the discrepancy of the radius ratios and to discriminate
a definitive atmosphere model for GJ1214b in the future,
further transit observations around K$_{\rm s}$
band would be especially important.
\end{abstract}

\section{Introduction}

Among several hundreds of discovered extrasolar planets,
transiting exoplanets can give us the most information about
properties of planets.
Transiting exoplanets are unique in that both the true mass
and the radius can be determined by measurements of
radial velocities and transit depths.
The mass-radius relation enables us to infer the internal structure
and bulk compositions of transiting exoplanets
(e.g., \cite{2007ApJ...665.1413V}).
One can also measure the angle between the stellar spin
axis and the planetary orbital axis in the sky projection
via the Rossiter-McLaughlin effect,
which provides insights of planetary migration histories
(e.g., \cite{2005ApJ...631.1215W, 2007PASJ...59..763N,
2008A&A...488..763H, 2009PASJ...61L..35N,
2009ApJ...703L..99W, 2011A&A...534L...6T,
2011ApJ...742...69H, 2012ApJ...757...18A}).
Another great merit of transiting planets is that they also provide
opportunities to probe atmospheric compositions
by transit photometry or spectroscopy.
Theoretical predictions have shown that  transit depths depend on wavelengths
(e.g., \cite{2000ApJ...537..916S, 2010ApJ...716L..74M,2012ApJ...756..176H}),
reflecting atmospheric compositions and environments (e.g., haze/clouds).
Previous space-based and ground-based observations for transiting hot Jupiters
have indeed revealed that this methodology,
often referred to as transmission spectroscopy,
is useful to learn exoplanetary atmospheres
(e.g., \cite{2002ApJ...568..377C, 2004PASJ...56..655W,
2005PASJ...57..471N,
2008ApJ...673L..87R, 2008A&A...487..357S,
2008Natur.452..329S, 2009ApJ...699..478D, 2011MNRAS.416.1443S}).

Recent discoveries of transiting planets with smaller radii (i.e., a few Earth radii)
around M dwarfs have expanded targets of this methodology down
to terrestrial exoplanets. 
Transiting planets around M dwarfs are especially favorable for this kind of studies
since transit depths become relatively deeper due to smaller host stars' radii.
In this sense, GJ1214b discovered by \citet{2009Natur.462..891C}
is currently the most interesting target.
Numbers of previous observers reported transit depths of GJ1214b
in various wavelength bands
(e.g., \cite{2010Natur.468..669B, 2011ApJ...743...92B, 2011ApJ...736...78C,
2011ApJ...731L..40D, 2011ApJ...730...82C, 2011ApJ...736...12B,
2012ApJ...747...35B, 2012A&A...538A..46D}).
The previous observations have revealed that there are
two possible atmospheric models to explain the observed transit
depths:
a water-dominated (higher mean molecular weight / lower scale height)
atmosphere and
a hydrogen-dominated (lower mean molecular weight / higher scale height)
atmosphere.
Among the previous studies, supporting evidences for the hydrogen-dominated
atmosphere are deeper transit depths observed in K$_{\rm s}$ band
\citep{2011ApJ...736...78C,2012A&A...538A..46D} and
in g band \citep{2012A&A...538A..46D}.
The rest of observations have shown a fairly flat transmission spectrum
for GJ1214b through visible and infrared wavelengths,
which indicates the water-dominated atmosphere or the hydrogen-dominated
but with high-altitude clouds/haze atmosphere.
Thus the nature of this planetary atmosphere is still an open question.

One known problem of GJ1214 as a target of transit photometry is
possible stellar variability and starspots,
which may cause small time-dependent variations in transit depths.
Some of the previous studies estimated that such small variations
would correspond to the ratio of radii of the planet and star
$R_{\rm p}/R_{\rm s} \sim 0.001$ \citep{2011ApJ...730...82C,
2011ApJ...736...12B, 2012A&A...538A..46D}, which are indeed
comparable to or larger than the nominal observational uncertainties
reported in the previous publications.
This fact makes it difficult to compare the all observed transit depths
in different epochs.
For this reason, simultaneous multi-band transit photometry is
the most useful way to discriminate possible atmosphere models.
We thus focus on JHK$_{\rm s}$ simultaneous transit photometry
since the largest difference in transit depths is theoretically predicted
between JH bands and K$_{\rm s}$ band.
In this sense, our motivation is the same as that by
\citet{2011ApJ...736...78C}, who tried nearly simultaneous
J and K$_{\rm s}$ band transit photometry by rapidly switching
the two bandpass filters.
Interestingly, \citet{2011ApJ...736...78C} claimed that K$_{\rm s}$ band
transit depth is significantly (5$\sigma$) deeper than that in J band.
With sufficient precision, we can independently confirm or refute
the claim by \citet{2011ApJ...736...78C} via JHK$_{\rm s}$ simultaneous
transit photometry.

In this paper, we present a result of such an observation using
the SIRIUS camera onboard the IRSF 1.4~m telescope.
This is indeed the first high precision transit observation using IRSF/SIRIUS.
We summarize target properties and observation methods in section 2,
and describe our analysis methods in section 3. 
We report our analysis results in section 4,
and discuss its meaning in section 5.
Finally, we summarize this paper in section 6.

\section{IRSF SIRIUS Observation for GJ1214}

\subsection{Target Properties}

The host star GJ1214 is an M4.5 type star at 13 pc away from the Sun,
with a mass of $0.157\pm0.019$ $M_{\odot}$ and a radius of
$0.2110\pm0.0097$ $R_{\odot}$ \citep{2009Natur.462..891C.
The stellar effective temperature is $3026\pm130$ K and the log of
the stellar surface gravity is $4.991\pm0.029$ in CGS units \citep{2009Natur.462..891C}.}
The star is known to show stellar variability at $\sim$2\% level with a period of
$\sim$52.3~d (see e.g., \cite{2011ApJ...736...12B}) and
also known to have starspots with surface coverage variability at several
percent level \citep{2011ApJ...730...82C}.

The orbiting planet GJ1214b is categorized as a super-Earth
with a mass of $6.55\pm0.98$ $M_{\oplus}$ and 
a radius of $2.68\pm0.13$ $R_{\oplus}$ \citep{2009Natur.462..891C}.
The planet orbits around GJ1214 at a period of $\sim$1.58 d and
a semi-major axis of $\sim$0.0146 AU, where is slightly closer to
the host star than the star's habitable zone ($a_{\rm HZ} \sim 0.06$ AU).
Previous investigation of transit timing variations (TTV) has shown
no evidence of large TTV in this system \citep{2011ApJ...730...82C}.

\subsection{Observation Setup}

We observed a full transit of GJ1214b during 18:43--21:00 of
UT 2011 August 14 with Simultaneous Infrared Imager for
Unbiased Survey (SIRIUS: \cite{2003SPIE.4841..459N})
on the Infrared Survey Facility (IRSF) 1.4~m telescope.
The IRSF and the SIRIUS camera were constructed and has been
operated by Nagoya University, SAAO (South African Astronomical
Observatory) at Sutherland, South Africa, and
National Astronomical Observatory of Japan.
The SIRIUS camera utilizes two dichroic filters and three
1024$\times$1024 HgCdTe detectors, which
can obtain J ($1.25\mu {\rm m} \pm 0.085\mu {\rm m} $),
H ($1.63\mu {\rm m}  \pm 0.15\mu {\rm m} $),
K$_{\rm s}$ ($2.14\mu {\rm m}  \pm 0.16 \mu {\rm m} $)
band images simultaneously,
with a square field of view of 7.7' on a side
and a pixel scale of 0.45''.
To achieve higher photometric precision,
we defocused the telescope so that stars have doughnut-like
point spread function (PSF).
The typical size of the PSF was $\sim13$ pixels in radius.
We carefully set GJ1214 and bright comparison stars away
from bad pixels on the detectors.
The exposure time was 49~s and the typical dead time was 7.6~s
(duty cycle of 86.6\%).
As a result, we obtained 145 frames during the above time
and use them for subsequent analyses.

\subsection{Position-Locking Software}

For high precision infrared photometry, it is known that
target positions on the detectors should be stared rather than
dithered (see e.g., \cite{2011ApJ...736...78C}).
Since the IRSF does not equip an auto-guider,
target positions tend to move over tens of pixels
during observations and it was the biggest problem
for precise transit observations with the IRSF.
To avoid such changes of target positions,
we have developed and installed a customized position-locking
software which gives appropriate feedback to the telescope
when the centroid position of the target star moves
over 1 pixel on the detectors.
By using the position-locking software,
GJ1214's centroid positions on the three detectors
are kept within an rms of about 2 pixels in both x and y directions.

\begin{table}[tpb]
\caption{Photometric transit light curves of GJ1214b taken by IRSF/SIRIUS}
\begin{center}
\begin{tabular}{ccc}
\hline
Time [BJD$_{\rm TDB}$]  & Value & Error \\
\hline
\multicolumn{3}{c}{J band}\\
\hline
2455788.28330 & 0.99979  &  0.00124 \\
2455788.28396 & 1.00011  &  0.00124 \\
2455788.28462 & 1.00125  &  0.00124 \\
2455788.28527 & 1.00137  &  0.00125 \\
2455788.28593 & 1.00015  &  0.00125 \\
\hline
\multicolumn{3}{l}{\hbox to 0pt{\parbox{80mm}{\footnotesize
$^{*}$ All data are presented in the electric table.
}\hss}}
\end{tabular}
\end{center}
\end{table}

\section{Data Reduction and Light Curve Analysis}

Primary data reduction is carried out with a pipeline for the
SIRIUS\footnote{http://www.z.phys.nagoya-u.ac.jp/\~{}nakajima/sirius/software/software.html},
including a correction for non-linearity,
dark subtraction, and flat fielding.
We note that the detectors of SIRIUS saturate at 32000 ADU (J band detector),
33000 ADU (H band detector), and 25000 ADU (K$_{\rm s}$ band detector) respectively
under the Correlated Double-Sampling (CDS) readout mode,
and keep a good linearity ($\le$1\%) up to $\sim$10000 ADU.
The above non-linearity correction enables us to work
up to $\sim$25000 ADU with $\le$1\% linearity.
The correction is necessary to reduce possible systematic errors
stemmed from non-linearity in fractional light curves down to $\le$0.1\%.
This is especially important for H and K$_{\rm s}$ bands where
sky background levels are high.

Aperture photometry is followed using a customized aperture photometry
pipeline \citep{2011PASJ...63..287F}, with a constant aperture-size mode
where a same aperture size is applied for all images.
Sky background levels on the observing night were typically $\sim1000$ ADU,
$\sim7000$ ADU, and $\sim6000$ ADU, for JHK$_{\rm s}$ bands, respectively.
Maximum counts in the frames (including sky background) are below 25000 ADU,
where the non-linearity correction can work.

Time stamps of observations are recorded in the
FITS headers in units of Modified Julian Day (MJD) based on
Coordinated Universal Time (UTC).
We convert the time system to Barycentric Julian Day (BJD)
based on Barycentric Dynamical Time (TDB) using
the code by \citet{2010PASP..122..935E}.

We then select appropriate comparison stars and aperture radii
for each band so that light curves have the smallest rms
at out-of-transit (OOT) phase as follows.
There are three possible comparison stars within the field of view of SIRIUS:
one is almost the same brightness as GJ1214 and the others are 1-2 magnitude fainter
than GJ1214.
First, we create fractional light curves by dividing fluxes of
GJ1214 by sum of all combinations of comparison stars' fluxes
with changing the aperture radius from 13 pixels to 30 pixels
in increments of 1 pixel.
All the fractional light curves $F_{\rm obs}$ exhibit systematic trends
(0.1\% level and apparently linear) at OOT phase.
The systematic trends could arise due to
slow variability in the brightness of GJ1214 itself or comparison stars,
changing airmass, or position changes of the stars on the detectors,
and so on.
We then correct the systematic trends by the following expression,
$F_{\rm cor} = F_{\rm obs} \times 10^{-0.4\Delta m_{\rm cor}}$,
by assuming the correction factor
$\Delta m_{\rm cor} = k_0 + k_t t + k_z z + k_x dx + k_y dy$,
where $t$ is the time, $z$ is the airmass, $dx$ and $dy$ are
the differences of centroid positions in x and y directions,
and $k_0, k_t, k_z, k_x, k_y$ are free coefficients.
We eliminate outliers (2, 10, 1 frames for JHK$_{\rm s}$ bands, respectively)
with large centroid position changes over 10 pixels,
and determine optimal coefficients using OOT
data by the AMOEBA algorithm \citep{1992nrca.book.....P}
so that the corrected OOT data have the smallest rms.
In this process, we also rescale the photometric errors of the data
so that reduced $\chi^2$ become unity.
We consequently choose to use only the brightest comparison star
(the same one for all bands)
and aperture radii of 15, 16, 15 pixels for JHK$_{\rm s}$ bands,
which give the smallest rms of the data as 0.00124, 0.00125, 0.00155
for JHK$_{\rm s}$ bands, respectively.
We note that including the other two comparison stars or
using different aperture radii have less impact on our subsequent conclusion,
but with slightly larger errors.
The data of the corrected transit light curves are presented in table~1.

For modeling the corrected transit light curves,
we follow the procedures that were adopted in previous
studies in litelature
\citep{2011ApJ...743...92B, 2011ApJ...736...78C,
2012A&A...538A..46D, 2012ApJ...747...35B}
so as to compare our transit depths with previous ones.
We fix the orbital inclination $i$ to 88.94$^{\circ}$ and
the orbital distance in units of the stellar radius $a / R_{\rm s}$
to 14.9749, which were determined by \citet{2010Natur.468..669B}.
We also fix the orbital period of GJ1214b
($P = 1.58040481$ d) and the mid-transit time
($T_{\rm c,0} = 2454966.525123$ BJD$_{\rm TDB}$),
determined by \citet{2011ApJ...743...92B}.
This assumption is justified by the fact that there is no evidence of
large TTV  \citep{2011ApJ...730...82C}.
We use the quadratic limb-darkening law,
$I(\mu) = 1 - u_1 (1-\mu) - u_2 (1-\mu)^2$,
where $I$ is the intensity and
$\mu$ is the cosine of the angle between
the line of sight and the line from the stellar center
to the position of the stellar surface.
We adopt empirical quadratic limb-darkening coefficients
($u_{\rm 1,J}=0.088$,
$u_{\rm 2,J}=0.404$,
$u_{\rm 1,H}=0.076$,
$u_{\rm 2,H}=0.407$,
$u_{\rm 1,Ks}=0.048$,
$u_{\rm 2,Ks}=0.350$)
for JHK$_{\rm s}$ bands given by \citet{2011A&A...529A..75C},
assuming stellar effective temperature $T_{\rm eff}=3000$ K
and log of the stellar surface gravity $\log g=5.0$.
We note that these assumptions on $T_{\rm eff}$ and $\log g$
are the same as \citet{2011ApJ...736...78C} for
J and K$_{\rm s}$ bands, although \citet{2011ApJ...736...78C}
used the non-linear limb-darkening law.

We fit the JHK$_{\rm s}$ transit light curves simultaneously
using the analytic formula given by \citet{2009ApJ...690....1O},
which is equivalent with \citet{2002ApJ...580L.171M} when using
the quadratic limb-darkening law.
The four free parameters are the difference of the observed mid-transit time 
from the predicted mid-transit time $\Delta T_c$ for this epoch (epoch $E=520$
from $T_{\rm c,0}$) and
the radius ratios of the planet and star $R_{\rm p}/R_{\rm s}$
for JHK$_{\rm s}$ bands.
In addition to the main analysis,
we also test some other cases with different conditions as follows:
(1) $T_{\rm c}$ for each band to be free,
(2) $u_2$ for each band to be free, and
(3) $i$ and $a/R_{\rm s}$ to be free,
to assess the validness of assumptions of fixed parameters above.
We employ the Markov Chain Monte Carlo (MCMC) method to
estimate values and uncertainties of the free parameters,
following the analysis by \citet{2011ApJ...736...78C}.
We create 5 chains of 5,000,000 points, then trim the first
100,000 points from each chain.
The acceptance ratios of jumping for the chains are set to about 25\%,
and we check that they pass the \citet{Gelman92} test.
We define $1\sigma$ statistical errors by the range of parameters between
15.87\% and 84.13\% of merged posterior distributions.

\begin{figure}[pthb]
 \begin{center}
  \FigureFile(85mm,85mm){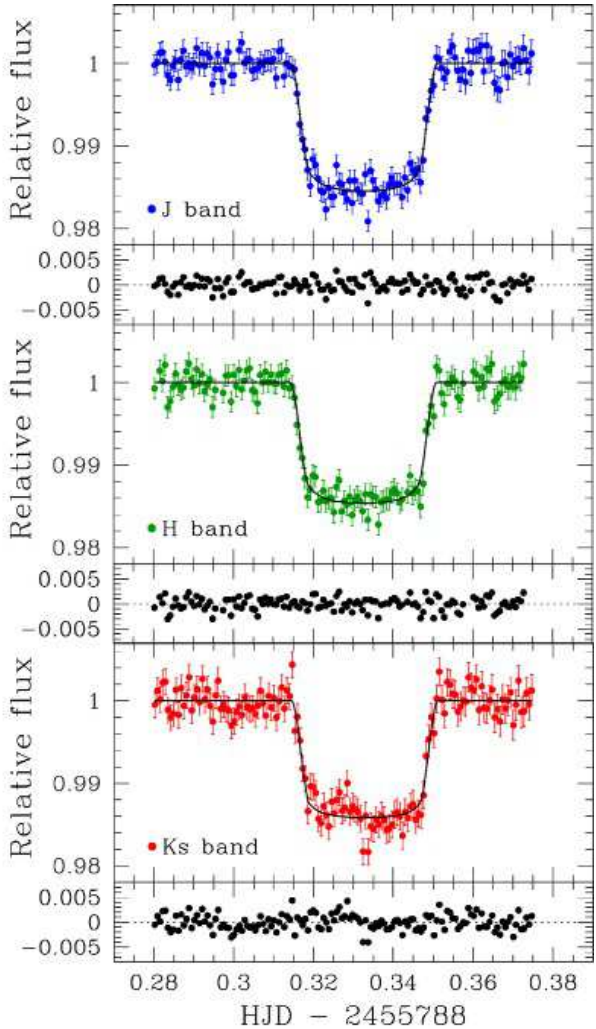}
 \end{center}
  \caption{Transit light curves with the best-fit models in J (top panel), H (middle panel),
  K$_{\rm s}$ (bottom panel) bands, observed simultaneously by the IRSF/SIRIUS.
  Each lower panel shows residuals between the observed
  values and the best-fit models.}
\end{figure}

\begin{figure}[pthb]
 \begin{center}
  \FigureFile(85mm,85mm){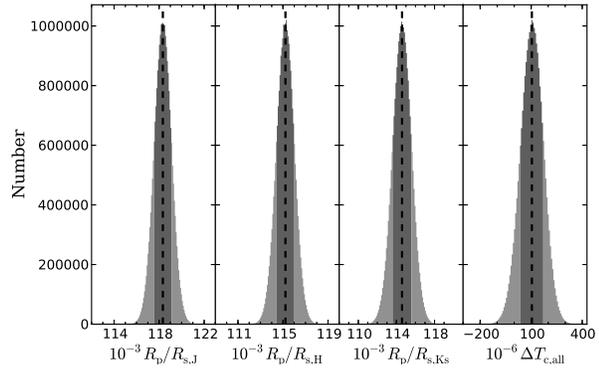}
 \end{center}
  \caption{MCMC histograms of the free parameters. The binning size is set to
  a hundredth part of the plotting region. The dashed vertical line
  in each panel indicates the mean value. The thick zones represent the 1$\sigma$
  statistical errors shown in table~2.}
\end{figure}

\begin{table*}[pthb]
\caption{Mean values and errors of the parameters based on the MCMC analyses.}
\begin{center}
\begin{tabular}{l|cc|cc|cc|cc}
\hline
\hline
Parameter & Value & Error$^{\dagger}$  & Value & Error & Value & Error & Value & Error \\
 & \multicolumn{2}{c}{main} & \multicolumn{2}{|c}{case (1)} &
 \multicolumn{2}{|c}{case (2)} & \multicolumn{2}{|c}{case (3)} \\
\hline
$R_{\rm p}/R_{\rm s}\!\,_{\rm ,J}$
&  0.11833 & $\pm 0.00077 $
&  0.11833 & $\pm 0.00077 $
&  0.11867 & $\pm 0.00095 $
&  0.11800 & $\pm 0.00083 $\\
$R_{\rm p}/R_{\rm s}\!\,_{\rm ,H}$
&  0.11522 & $\pm 0.00079 $
&  0.11519 & $\pm 0.00079 $
&  0.11668 & $\pm 0.00096 $
&  0.11492 & $\pm 0.00085 $\\
$R_{\rm p}/R_{\rm s}\!\,_{\rm ,Ks}$
&  0.11459 & $\pm 0.00099 $
&  0.11463 & $\pm 0.00099 $
&  0.11428 & $\pm 0.00121 $
&  0.11427 & $\pm 0.00101 $\\
$\Delta T_{\rm c,all}$ [s] 
&  0.000104 & $\pm 0.000067 $
&  -- & -- 
&  0.000107 & $\pm 0.000066 $
&  0.000102 & $\pm 0.000069 $\\
$\Delta T_{\rm c,J}$ [s] 
&  -- & -- 
&  0.000111 & $\pm 0.000111 $
&  -- & -- 
&  -- & -- \\
$\Delta T_{\rm c,H}$ [s]
&  -- & -- 
&  0.000005 & $\pm 0.000114 $
&  -- & -- 
&  -- & -- \\
$\Delta T_{\rm c,Ks}$ [s]
&  -- & -- 
&  0.000244 & $\pm 0.000134 $
&  -- & -- 
&  -- & -- \\
$u_{\rm 2,J}$ 
&  0.404 & fixed
&  0.404 & fixed 
&  0.345 & $\pm 0.095$
&  0.404 & fixed \\
$u_{\rm 2,H}$ 
&  0.407 & fixed
&  0.407 & fixed
&  0.145 & $\pm 0.103 $
&  0.407 & fixed \\
$u_{\rm 2,Ks}$ 
&  0.350 & fixed
&  0.350 & fixed
&  0.403 & $\pm 0.125 $
&  0.350 & fixed \\
$i$ [$^{\circ}$]
& 88.94 & fixed
& 88.94 & fixed
& 88.94 & fixed
& 89.13 & $^{+0.53}_{-0.44}$ \\
$a / R_{\rm s}$
& 14.9749 &  fixed
& 14.9749 &  fixed
& 14.9749 &  fixed
& 15.0468 &  $^{+0.3413}_{-0.4847}$ \\
\hline
\hline
\multicolumn{9}{l}{\hbox to 0pt{\parbox{160mm}{
$^{\dagger}$ The presented errors do not include systematic errors due to the fixed
parameters (see text).
}\hss}}
\end{tabular}
\end{center}
\end{table*}

\section{Results}

The mean values and 1$\sigma$ statistical errors based on the MCMC analyses
are summarized in table~2.
We note that the errors do not include systematic errors discussed
in the next section.
In the main analysis, we find $\Delta T_{\rm c} = 0.000104 \pm 0.000067$~s,
which is consistent with the predicted mid-transit time within 2$\sigma$,
validating the assumption to use the fixed period
determined by \citet{2011ApJ...743...92B}.
The derived radius ratios are
$R_{\rm p}/R_{\rm s,J} = 0.11833 \pm 0.00077$,
$R_{\rm p}/R_{\rm s,H} = 0.11522 \pm 0.00079$,
$R_{\rm p}/R_{\rm s,Ks} = 0.11459 \pm 0.00099$,
respectively.
Comparisons of our result with the previous studies are discussed in the next section.
Figure~1 plots the transit light curves with the models using the mean MCMC values,
and figure~2 shows histograms of the posterior distribution of the free parameters.

One may wonder why our errors are comparable with \citet{2011ApJ...736...78C},
even though we observed only one transit with the IRSF 1.4m telescope and
\citet{2011ApJ...736...78C} observed 3 transits with the CFHT 3.8m telescope.
The reasons are that \citet{2011ApJ...736...78C} switched between J and K$_{\rm s}$ band filters
during the transits and the 3.8m telescope was too large for the brightness of GJ1214.
Namely, \citet{2011ApJ...736...78C} employed 4 s exposure and the total duty cycle was
only 22\% for J and K$_{\rm s}$ bands, which is significantly lower than our duty cycle of 86.6\%.
Considering the above fact, it is reasonable that our errors are similar to those by \citet{2011ApJ...736...78C}.

From the case (1), we find that $\Delta T_{\rm c}$ for JHK$_{\rm s}$ bands
are all consistent with zero within $2\sigma$ and one another.
Also, the result shows that fitting $\Delta T_{\rm c}$ for JHK$_{\rm s}$ bands
at once or separately has less impact on radius ratios.

The case (2) illustrates a potential of systematic errors and
underestimated errors stemmed from the fixed limb-darkening coefficients.
The derived $u_2$ for J and K$_{\rm s}$ bands are well consistent
with the empirical values by \citet{2011A&A...529A..75C}, and then
the radius ratios in the same bands are also well accorded
with the main result.
While the derived $u_2$ in H band is $\sim$2.5$\sigma$ apart from
the empirical value, and the radius ratio is also different from the main result
by $\sim$1.5$\sigma$.
Although some systematic errors may be present, we consider that the assumption
of the adopted limb-darkening coefficients is reasonable for the mean values,
since the empirical values are in excellent agreement with the derived values in
J and K$_{\rm s}$ bands.
For this reason, we believe that the radius ratio in H band using the empirical
limb-darkening coefficients (the main case) is more reliable than that
with free $u_2$ (the case 2).
Another fact seen from the case (2) is that
the errors of the radius ratios are about 20\% larger than
the main case where the limb-darkening coefficients are fixed.
This fact is often pointed out in previous studies
(e.g., \cite{2008MNRAS.386.1644S}).
Thus we note that the quoted errors in the radius ratios with
the fixed limb-darkening coefficients may have 20\% additional errors.

Finally, we confirm that the orbital inclination ($i$)
and the orbital distance in units of the stellar radius ($a/R_{\rm s}$)
derived from our IRSF data are well consistent with
those by \citet{2010Natur.468..669B} via the analysis of case (3).
We also confirm the derived radius ratios are consistent with
the main result, however we notice that the errors are
about 10\% larger than the main result.
This fact also suggests that errors may be about 10\% underestimated
when using the fixed $i$ and $a/R_{\rm s}$.

\section{Discussions}

\subsection{Impacts of Systematic Errors}

In this section, we compare our radius ratios in JHK$_{\rm s}$ bands
with the previous studies and atmospheric models.
Before doing that, we should care about possible systematic errors.

First, as assessed by the test cases (2) and (3),
we note that our statistical errors may have additional 30\%
uncertainties due to the systematic errors stemmed from the fixed parameters
(20\% from fixing the limb-darkening coefficients, and
10\% from fixing the orbital inclination and the orbital distance).

Second, it is also necessary to consider impacts of stellar flux variability and stellar spots
on our results so as to compare radius ratios with previous studies
measured in different epochs and wavelengths.
\citet{2011A&A...526A..12D} presented an equation for estimating
a difference of radius ratio $\Delta (R_{\rm. p}/R_{\rm s})$
caused by stellar flux variability due to unocculted spots
(see equation 7 of \cite{2011A&A...526A..12D}).
The equation can be rewritten as
\begin{equation}
\Delta (R_{\rm p}/R_{\rm s}) \simeq  0.5\,\, \Delta f(\lambda)\,\, (R_{\rm p}/R_{\rm s}),
\end{equation}
where $\Delta f(\lambda)$ is stellar flux variability at wavelength $\lambda$.
\citet{2011ApJ...736...12B} reported stellar flux variability of $\sim2\%$ level
in the MEarth observing bandpass ($\sim780$nm), and 
if following the assumption used in \citet{2011ApJ...736...78C} that
GJ1214 ($T_{\rm eff} \sim 3000$K) has spots 500 K cooler,
then the stellar flux variabilities in JHK$_{\rm s}$ bands can translate into
$\sim1.5\%, \sim1.3\%, \sim1.0\%$, respectively \citep{2011ApJ...736...78C}.
Using the equation above, 
possible variability of radius ratios in JHK$_{\rm s}$ bands are
$\sim0.75\%, \sim0.65\%, \sim0.5\%$ of the observed radius ratios.
Thus we should take into account possible systematic differences of
$\Delta (R_{\rm p}/R_{\rm s,J})\sim0.00089$,
$\Delta (R_{\rm p}/R_{\rm s,H})\sim0.00075$,
$\Delta (R_{\rm p}/R_{\rm s,Ks})\sim0.00057$,
when comparing our radius ratios with other epochs' results.

Third, although we do not find any clear sign of spot occultation in our data, 
possible existence of unocculted spots is still a problem when comparing
radius ratios with theoretical atmospheric models, since their impact depends
not only on wavelength but also on spot properties
(coverage, temperature, and so on)
and GJ1214's spot properties are still unclear.
\citet{2012A&A...538A..46D} presented corrections for radius ratios
assuming several cases of different spot coverage.
According to their study, the corrections vary greatly between optical and
infrared wavelength ($\Delta (R_{\rm p}/R_{\rm s})\sim0.003$ if spot coverage
is 5\%), but corrections in JHK$_{\rm s}$ bands are very similar
(see figure 7 of \cite{2012A&A...538A..46D}).
Thus we neglect the corrections for the current study since our data are
only in JHK$_{\rm s}$ bands, although we should take care when comparing
our radius ratios with those in optical bands.

\begin{figure}[pthb]
 \begin{center}
  \FigureFile(85mm,85mm){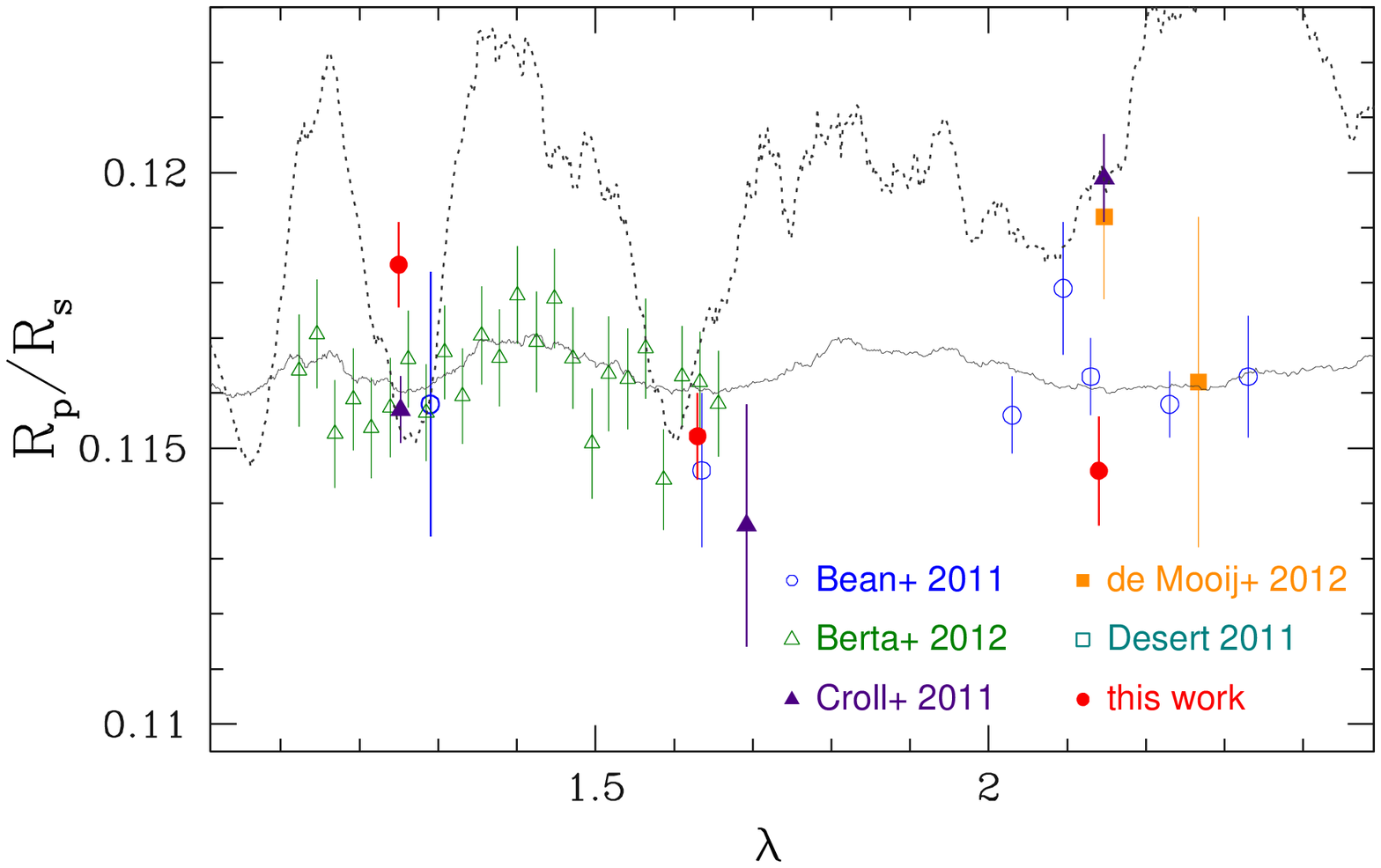}
  \FigureFile(85mm,85mm){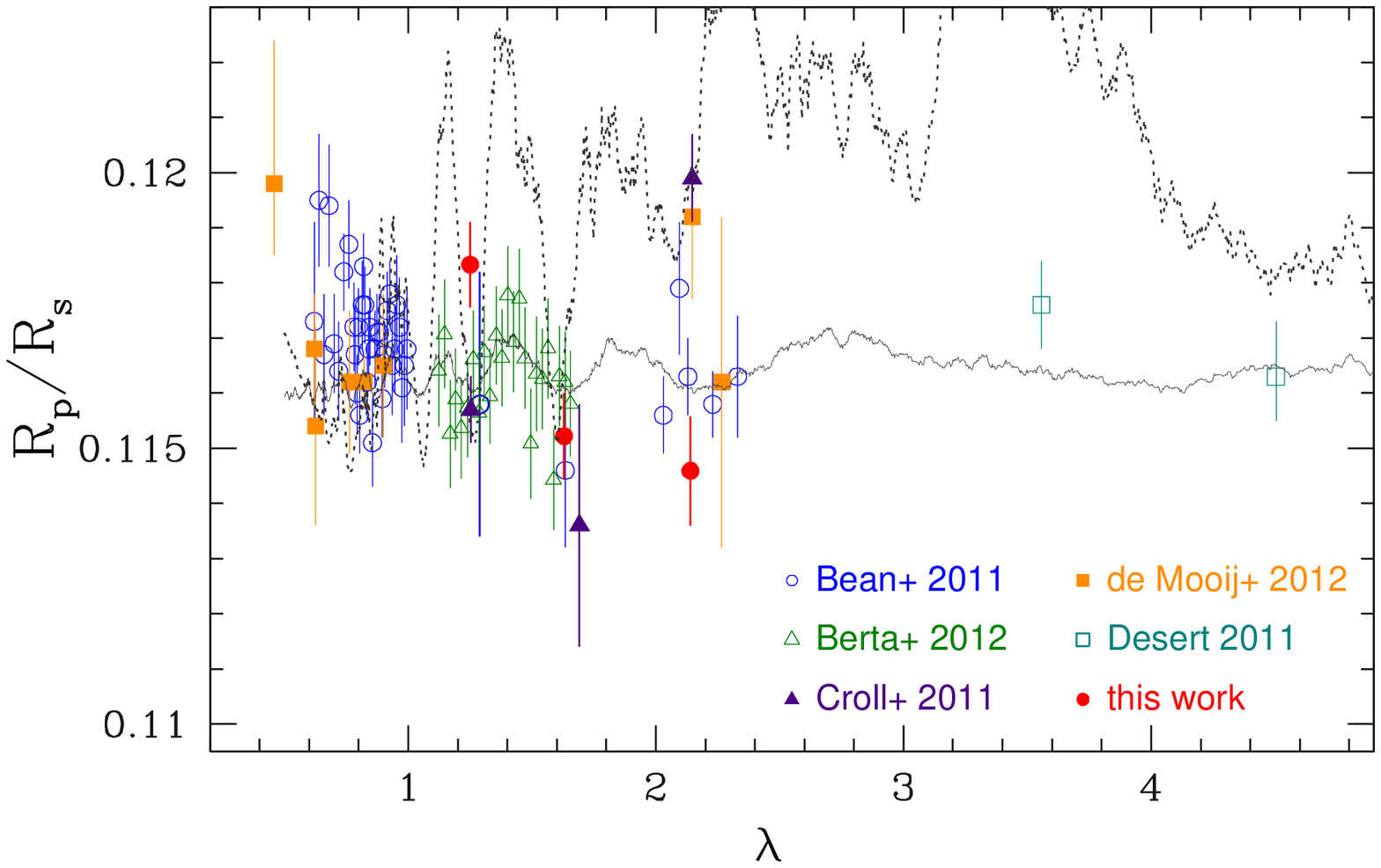}
 \end{center}
  \caption{Comparisons of our results with the previous observational studies
  and the cloudless atmospheric models by \citet{2010ApJ...716L..74M}.
  The gray solid line represents the water-dominated (vapor) atmosphere model,
  and the gray dotted line does the hydrogen-dominated (Solar composition) model.
  The upper figure shows 1-2.5 $\mu$m wavelength region and the lower one
  does 0.6-4.5 $\mu$m region.}
\end{figure}

\subsection{Comparisons with the Previous Studies}

Figure~3 plots our observed radius ratios with those in the previous studies
\citep{2011ApJ...743...92B,2011ApJ...736...78C,2011ApJ...731L..40D,
2012ApJ...747...35B,2012A&A...538A..46D}
and the two atmospheric models (i.e., hydrogen-dominated and
water-dominated atmospheres) by \citet{2010ApJ...716L..74M}.
Note that the values and errors of the previous studies are reported ones
without any corrections, and
the errors of our data are based on the MCMC analysis
(30\% additional errors are not included, since previous observers did not
take into account such systematic errors).

In summary, our results support a featureless transmission spectrum with
a shallower (without a deeper) transit in K$_{\rm s}$ band,
which favors a water-dominated atmosphere rather than
a cloudless hydrogen-dominated atmosphere by \citet{2010ApJ...716L..74M}.
A similar flat transmission spectrum can be expected
if the atmosphere is covered by optically thick high-altitude
clouds or haze (see e.g., \cite{2011ApJ...743...92B, 2012ApJ...747...35B}).

A similar conclusion was also reported by \citet{2011ApJ...743...92B},
who observed two transits of GJ1214b using
the MMIRS instrument of the Magellan telescope by the multi-object
spectroscopy approach. They obtained radius ratios of GJ1214b
in JHK bands (almost the same bandpass with our bands) as
$R_{\rm p}/R_{\rm s,J} = 0.1158 \pm 0.0024$,
$R_{\rm p}/R_{\rm s,H} = 0.1146 \pm 0.0014$,
$R_{\rm p}/R_{\rm s,K} = 0.1158 \pm 0.0006$, respectively.
Thus our radius ratios in JHK$_{\rm s}$ bands are all consistent with
\citet{2011ApJ...743...92B} within a square-root of sum of squares
of both 1$\sigma$ errors.
Our radius ratios in J and H bands are also in agreement with
\citet{2012ApJ...747...35B}, who conducted spectro-photometry
in the bandpass between 1.1 and 1.7 $\mu$m using
the Wide Field Camera 3 (WFC3) onboard the Hubble Space Telescope.

On the other hand, our radius ratio in K$_{\rm s}$ band appears
inconsistent with the previous measurements in the same band
by \citet{2011ApJ...736...78C} ($R_{\rm p}/R_{\rm s,Ks} = 0.1199 \pm 0.0008$)
and \citet{2012A&A...538A..46D} ($R_{\rm p}/R_{\rm s,Ks} = 0.1189 \pm 0.0015$).
Especially, our result
($R_{\rm p}/R_{\rm s,Ks} = 0.1146 \pm 0.0010$) differs about
4$\sigma$ (1$\sigma$ here is a square-root of sum of squares of both
1$\sigma$ errors) from that by \citet{2011ApJ...736...78C}.
The discrepancy cannot be explained by stellar flux variability in K$_{\rm s}$ band
($\Delta (R_{\rm p}/R_{\rm s,Ks})\sim0.00057$)
nor 30\% additional systematic errors
due to the fixed parameters.
Thus we do not confirm a deeper transit in K$_{\rm s}$ band and
we have no clear explanation for this discrepancy at this point in time.
Further transit observations in K$_{\rm s}$ band would be necessary
to solve this enigma.

\subsection{Possible Origins of Water-Dominated Atmospheres}

The investigation of planetary atmospheres provides a crucial clue
to learn the bulk composition of super-Earths like GJ1214b,
which also gives important constraints on formation processes
of such planets.
Only with its measured mass and radius, we are unable to distinguish
whether GJ1214b is a rocky planet with a thick H/He atmosphere or
a water-dominated atmosphere \citep{2010ApJ...716.1208R,2011ApJ...733....2N}.
Transmission spectroscopy is thus an important key to discriminate
the atmospheric nature of transiting planets.
As discussed earlier, our finding as well as some of the previous studies
(e.g., \cite{2011ApJ...743...92B,2011ApJ...731L..40D,2012ApJ...747...35B})
support a possibility of a water-dominated atmosphere for GJ1214b.
Then, how can such atmospheres form?

According to detailed modeling of the internal structure by \citet{2011ApJ...733....2N},
however, two-layer models (a water envelope on top of a rock core) which are
consistent with GJ1214's age would have unreasonably
high water-to-rock ratio.
To fix this problem, a small amount of H/He,
which accounts for several percent of the planet's total mass,
must be incorporated in the water envelope (see Figure~4 of \cite{2011ApJ...733....2N}).
In other words, GJ1214b is a Uranus/Neptune-like planet whose atmosphere
is heavily polluted by water.

Such properties of GJ1214b are qualitatively consistent with recent
theories of planet formation.
GJ1214b is orbiting close to its host star, in contrast to Uranus and Neptune.
Its current location was too hot for icy building blocks to exist
in the proto-planetary disk where the planet was born.
Therefore, provided GJ1214b contains a significant amount of water,
it is reasonable to think that the planet has experienced orbital migration.

One possible mechanism for migration is angular-momentum exchange
with the proto-planetary disk (i.e., the type-I migration; \cite{1986Icar...67..164W}).
Recent $N$-body simulations of planetary accretion with the effect of
the type-I migration have demonstrated that
water-rich proto-planets of super-Earth-size can form
in the vicinity of the central star \citep{2009ApJ...699..824O}.
As the proto-planets are embedded in the disk,
they also capture the ambient H/He disk gas naturally.
A proto-planet with mass similar to that of GJ1214b can gain H/He atmosphere
comparable in mass with that predicted by \citet{2011ApJ...733....2N},
although it depends on values of several parameters \citep{2012ApJ...753...66I}.
Then, bombardments of smaller planetary embryos and/or other proto-planets
rich in water can heavily pollute the H/He atmosphere by water
during and after the migration.

Another possibility is that a Uranus/Neptune-like planet that had formed
beyond the snow line has migrated by gravitational scattering or
by long-term interaction with outer gravitationally perturbing objects.
Beyond the snow line, icy planetesimals are sufficiently
available for polluting H/He atmospheres.
Indeed, the metallicity of the atmospheres of Uranus and Neptune are
known to be significantly super-solar (\cite{1995netr.conf..109H}),
which suggest such pollutions.

While both scenarios are to be quantitatively verified, the both scenarios suggest
that there are other planets yet undetected in the GJ1214b system,
although no TTV sign has been found (e.g., \cite{2011ApJ...730...82C}).
Future search for additional planets with RV measurements
(possibly using brighter NIR region) is expected to verify the above scenarios.
Moreover, a measurement of the spin-orbit alignment angle of GJ1214b via
the Rossiter-McLaughlin effect would be very helpful in
knowing which of the above two scenarios is true.

\section{Summary}

We conducted high precision JHK$_{\rm s}$ simultaneous transit photometry
for GJ1214b using the SIRIUS camera on the IRSF at South Africa.
It was the first high precision transit observation using IRSF/SIRIUS.
In this process, we have demonstrated that high precision NIR transit
photometry is possible even without an auto-guider by
the position-locking software installed on the telescope.
We have found a featureless transmission spectrum in the observed bands,
which suggests a water-dominated atmosphere or an atmosphere
with extensive high-altitude clouds/haze.
Although the observed radius ratios are well consistent with the results by
\citet{2011ApJ...743...92B} and \citet{2012ApJ...747...35B},
our result in K$_{\rm s}$ band is unaccountably inconsistent with
the previous studies in the same band by
\citet{2011ApJ...736...78C}.
We have no good explanation for this discrepancy at this point.
To solve the discrepancy and to distinguish a definitive atmosphere model
for this planet, further transit observations around K$_{\rm s}$ band would
be especially important.
In addition, it will be also interesting to observe transits in blue optical bands
in order to ascertain the Rayleigh scattering in the atmosphere of GJ1214b.
With such additional observations, we will be able to learn the true
nature of the atmosphere of GJ1214b, and
discriminating the true atmospheric nature would also give us
useful insights on the formation process of this planet.

\bigskip

We acknowledge Eliza Miller-Ricci Kempton for kindly providing their
atmospheric models.
We thank Eric Gaidos, Teruyuki Hirano, Masayuki Kuzuhara, Hiroshi Ohnuki, and 
Yasuhiro Takahashi for fruitful discussions.
We acknowledge a support by NINS Program for Cross-Disciplinary Study.
N.N. is supported by NAOJ Fellowship and by the JSPS Grant-in-Aid for
Research Activity Start-up No. 23840046.
The IRSF project was financially supported by the Sumitomo foundation
and Grants-in-Aid for Scientific Research on Priority Areas (A) (Nos.
10147207 and 10147214) from the Ministry of Education, Culture,
Sports, Science and Technology (MEXT).
The operation of IRSF is supported by Joint Development Research of
National Astronomical Observatory of Japan, and  Optical Near-Infrared
Astronomy Inter-University Cooperation Program, funded by the MEXT.
This work is partly supported by a Grant-in-Aid for
Specially Promoted Research, No. 22000005 from the MEXT
and by the Mitsubishi Foundation.



\end{document}